\documentclass{amsart}
\usepackage{amsmath,amssymb,amsfonts}
\usepackage{graphicx}
\usepackage{amsrefs}

\def\sign{\mathrm{sign\,}}

\def\const{\mathrm{const}}

\def\T{\mathrm{T}}

\def\E{\mathbf{E}}

\newtheorem{theorem}{Theorem}
\newtheorem{proposition}{Proposition}

\newtheorem{remark}{Remark}

\date{}

\begin{document}

\author{Vladimir Dragovi\'c}
\address{Mathematical Institute SANU, Kneza Mihaila 36, Belgrade,
Serbia\newline\indent Mathematical Physics Group, University of
Lisbon, Portugal} \email{vladad@mi.sanu.ac.rs}

\author{Milena Radnovi\'c}
\address{Mathematical Institute SANU, Kneza Mihaila 36, Belgrade, Serbia}
\email{milena@mi.sanu.ac.rs}

\title{Bifurcations of Liouville Tori in Elliptical Billiards}

\maketitle

\begin{abstract}
A detailed description of topology of integrable billiard systems
is given. For elliptical billiards and geodesic billiards on
ellipsoid, the corresponding Fomenko graphs are constructed.
\end{abstract}

\tableofcontents

\section{Introduction}

An important and geometrically significant class of dynamical
systems consists of billiards with ellipsoidal boundary in $\E^d$.
By Chasles theorem, any line in the $d$-dimensional space is tangent
to exactly $d-1$ quadrics from a given confocal family. These $d-1$
caustics, confocal to the boundary of the billiard, are fixed for
each segment of a given trajectory \cite{ArnoldMMM}. Existence of
caustics is, in fact, a geometrical manifestation of integrals of
motion. Due to them, we easily see that elliptical billiard is an
completely integrable Hamiltonian system.

\smallskip

Famous classical example of integrable systems is geodesic flow on
an ellipsoid $\mathcal E$ in $\E^d$ \cite{JacobiGW}. This system is
closely related with billiards. In one way, the geodesic flow tends
to the billiard flow inside an ellipsoid in $\E^{d-1}$ when the
smallest axis of $\mathcal E$ tends to zero. On the other hand, the
geodesic flow on $\mathcal E$ is a limit of the billiard motion
inside $\mathcal E$, when one of the caustics is a confocal
ellipsoid tending to the boundary. Geodesic flow on ellipsoid also
gives rise to another interesting class of billiard systems:
billiards on the ellipsoid, where the boundary is determined by the
intersection of the ellipsoid with a confocal quadric. Trajectories
of such a billiard are composed of geodesic segments.

\smallskip

Billiard systems within and geodesic flows on ellipsoids have been
extensively studied from XIXth century on: see classical works
\cite{Cayley1854,DarbouxSUR,JacobiGW,LebCONIQUES,Poncelet1822} and
more recent ones
\cite{AbendaFed2004,Audin1994,CCS1993,CS1989,DragRadn2006,DragRadn2008,GrifHar1977,GrifHar1978,Knorr1980,Mo1980,MoVes1991,Previato2002,WDullin2002}
with the references therein. From the non-vanishing interest, one
may notice that this topic is far from being exhausted.

\smallskip

The object of this paper is to give topological description of such
systems using Fomenko graphs. The detailed description of this kind
of topological classification of integrable systems can be found in
\cite{BolFomBOOK,BMF1990,BO2006} and references therein, while a
very concise summary for a reader not acquainted with it is given in
the appendix of this paper. Although plane elliptical billiards are
quite well known, we may see that such a description will give a new
and exciting insight into its properties -- note, for example,
appearance of a non-orientable periodic trajectory along one of the
axes of the billiard boundary in Proposition \ref{prop:ellipse.hyp},
see Figure \ref{fig:bilijar3}.

\smallskip

In the book \cite{BolFomBOOK}, one may find a large list of Fomenko
graphs for known integrable systems, such as integrable cases of
rigid body motion and integrable geodesic flows on surfaces. This
paper provides an additional set of such examples. Some other
examples, connected with near-integrable dynamics can be found in
\cite{SRK2005} while bifurcations of Liouville foliations in a
certain class of integrable systems with two degrees of freedom,
were classified using Fomenko graphs in \cite{RadnRK2008}.

\smallskip

Let us note that topological properties of elliptical billiards,
viewed as discrete dynamical systems, have been recently studied in
\cite{WDullin2002}. However, in this paper, we consider billiards as
continuous systems, which enables us to use tools developed by
Fomenko and his school.

\smallskip

This article is organized as follows. In Section
\ref{sec:isoenergy}, we begin with introductory remarks on
isoenergy surfaces of billiard systems. Section \ref{sec:ravan}
contains topological description of several examples of plane
billiards, with the boundary composed of confocal conics, together
with their representation via Fomenko graphs. In Section
\ref{sec:na.elipsoidu}, topological description of the billiard
flow with the quadratic boundary on the ellipsoid in $\E^3$ is
given. We conclude this section by demonstrating the Liouville
equivalence of the billiard motion within an ellipsoid on any
Liouville surface. In Section \ref{sec:vise.dimenzija}, the
topological structure of the billiard inside ellipsoid in $\E^3$
is described. The Appendix contains a summary on Fomenko graphs.

\section{Isoenergy Surfaces of Billiard
Systems}\label{sec:isoenergy}

Let $\mathcal M^n$ be an $n$-dimensional Riemann manifold, and
$\Omega\subset\mathcal M$ a domain with the boundary composed of
several smooth hypersurfaces. The billiard \cite{KozTrBIL} inside
$\Omega$ is a dynamical system where a material point of the unit
mass is freely moving inside the domain and obeying the reflection
law at the boundary, i.e.\ having congruent impact and reflection
angles with the space tangent to the boundary at any bouncing point.
It is also assumed that the reflection is absolutely elastic, i.e.\
that the speed of the material point does not change before and
after impacts.

\smallskip

It is important to remark that the change of the total energy of
the system, only the intensity of the velocity vector of the point
will be changed while the trajectories will be the same at each
energy level. That is why, for the complete analysis of the
billiard system, we may fix the speed and investigate only one
energy level.

\smallskip

The isoenergy space for the billiard system inside $\Omega$ is:
$$
\displaylines{
 \mathcal B= \{\ (x,v) \ \mid\ x\in\Omega,\ v\in\T_x\mathcal M,\ |v|=1\ \}/\sim\,,
 \cr
 (x,u)\sim(x,v)\quad\Leftrightarrow\quad x\in\partial\Omega\ \ \text{and}\ \ u-v\perp\T_x\partial\Omega. }
$$

Although the smoothness of motion of the billiard particle is
violated at the boundary, let us note that, unexpectedly, the
isoenergy space will be smooth, assuming that the billiard
boundary is smooth. Even more, this manifold is smooth also when
boundary is only composed of a few smooth parts, if the billiard
reflection can be continuously defined at the angle points. This
will be always the case in our examples, when the boundary is
placed on a few confocal conics, since they are orthogonal to each
other. Thus, the reflection at the intersection points is defined
as follows: the velocity vector $v$ after the reflection at the
non-smooth point of the boundary, changes to $-v$.

\section{Plane Elliptical Billiards}\label{sec:ravan}

An important example of integrable billiard system is the billiard
within an ellipse in $\E^2$. The integrability of such a billiard
motion is due to the nice and elementary fact: each segment of a
trajectory is tangent to the same conic confocal with the boundary
\cite{ArnoldMMM, BergerGeometry}. Moreover, the boundary of any
integrable billiard in a plane domain is composed of segments of
several confocal conics \cite{Bolotin1990}.

\smallskip

In this section, we will describe the topology of the plane
billiards with the elliptic boundary. We will use the Fomenko graphs
\cite{BMF1990} to represent the isoenergy surfaces.

\smallskip

\smallskip

Suppose that the ellipse in the plane is given by:
$$
\mathcal E\ :\ \frac{x^2}a + \frac{y^2}b = 1, \quad a>b>0.
$$
The family of conics confocal with $\mathcal E$ is:
$$
\mathcal C_{\mu}\ :\ \frac{x^2}{a-\mu} + \frac{y^2}{b-\mu} = 1,
\quad \mu\in\mathbf R.
$$

We are going to consider billiard systems inside a bounded domain
$\Omega$ in $\E^2$, whose boundary is a union of arcs of several
confocal conics. As in Introduction, we denote its isoenergy
manifold by $\mathcal B$.

\begin{proposition}\label{prop:bilijar1}
The isoenergy manifold corresponding to the billiard system within
an ellipse in $\E^2$ is represented by the Fomenko graph on Figure
\ref{fig:fom1}.
\begin{figure}[h]
\centering
\includegraphics[width=3cm,height=4cm]{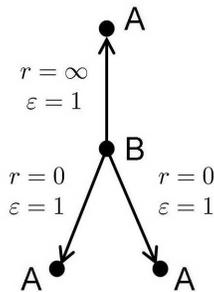}
{\caption{Fomenko graph corresponding to the billiard within an
ellipse}\label{fig:fom1}}
\end{figure}

The rotation functions corresponding to the edges of the graph are
monotonous, and the limits of these functions on the lower edges
are equal to $\infty$ approaching to $\mathbf A$-atom and $2$
approaching to $\mathbf B$-atom; on the upper edge the limit is
$\infty$ approaching to $\mathbf A$-atom and $1$ approaching to
$\mathbf B$-atom.
\end{proposition}

\begin{proof}
The isoenergy manifold $\mathcal B$ is the solid torus with the
identification on $\sim$ on the boundary. This identification
glues pairs of points, while points contained on two curves --
representing flows in positive and negative directions along the
boundary of the billiard table, are not glued with any other. The
torus with the two curves on the boundary is shown on Figure
\ref{fig:torus1}.
\begin{figure}[h]
\centering
\includegraphics[width=5cm,height=4cm]{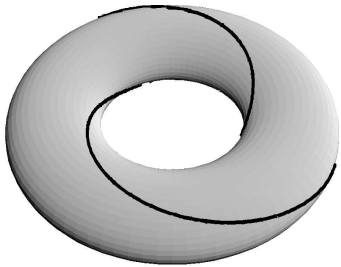}
\caption{The isoenergy manifold for the billiard system within an
ellipse}\label{fig:torus1}
\end{figure}
These two curves correspond to the limit flow with the caustic
$\mathcal C_0=\mathcal E$, and they are represented with the lower
$\mathbf A$-atoms on the graph. For the parameter $0<\mu<b$, i.e.
when the caustic is an ellipse, we have two Liouville tori over
each value of $\mu$ -- one for each direction of rotation.

\smallskip

Let us describe the level set $\mu=b$. This level set contains
exactly those billiard trajectories that pass through foci of the
ellipse. Among them, there is one periodic trajectory -- the
motion along $x$-axis, while others have the well-known property
that their segments alternately pass through left and right focus
of the ellipse. These trajectories can be naturally divided into
$2$ classes -- one is composed of the trajectories where the
particle is moving upward through the left focus and downward
through the right one, and the second contains the trajectories
with the reverse property. Note these two classes by $\mathcal
S_1$ and $\mathcal S_2$ respectively. All their trajectories are
homoclinically tending to the $x$-axis, thus $\mathcal S_1$ and
$\mathcal S_2$ are separatrices. Note that the periodic trajectory
is orientable, so this level set corresponds to the
 $\mathbf B$-atom.

\smallskip

Note that $\mathcal S_1$ (resp.\ $\mathcal S_2$) is the limit set
of the family of Liouville tori corresponding to the flow with
elliptic caustic in the clockwise (resp.\ counterclockwise)
direction.

\smallskip

When $b<\mu<a$, i.e.\ when the caustic is a hyperbola, there is
only one torus for each value. Finally, to the $\mu=a$, the
periodic motion along $y$-axis takes place, and it is represented
by the upper $\mathbf A$-atom.
\end{proof}

In the similar way, we can obtain the following:

\begin{proposition}\label{prop:ell.ell}
The isoenergy manifold corresponding to the billiard system in the
domain limited with two confocal ellipses in $\E^2$ is represented
by the Fomenko graph on Figure \ref{fig:fom2}.

\begin{figure}[h]
\centering
\includegraphics[width=5cm,height=4cm]{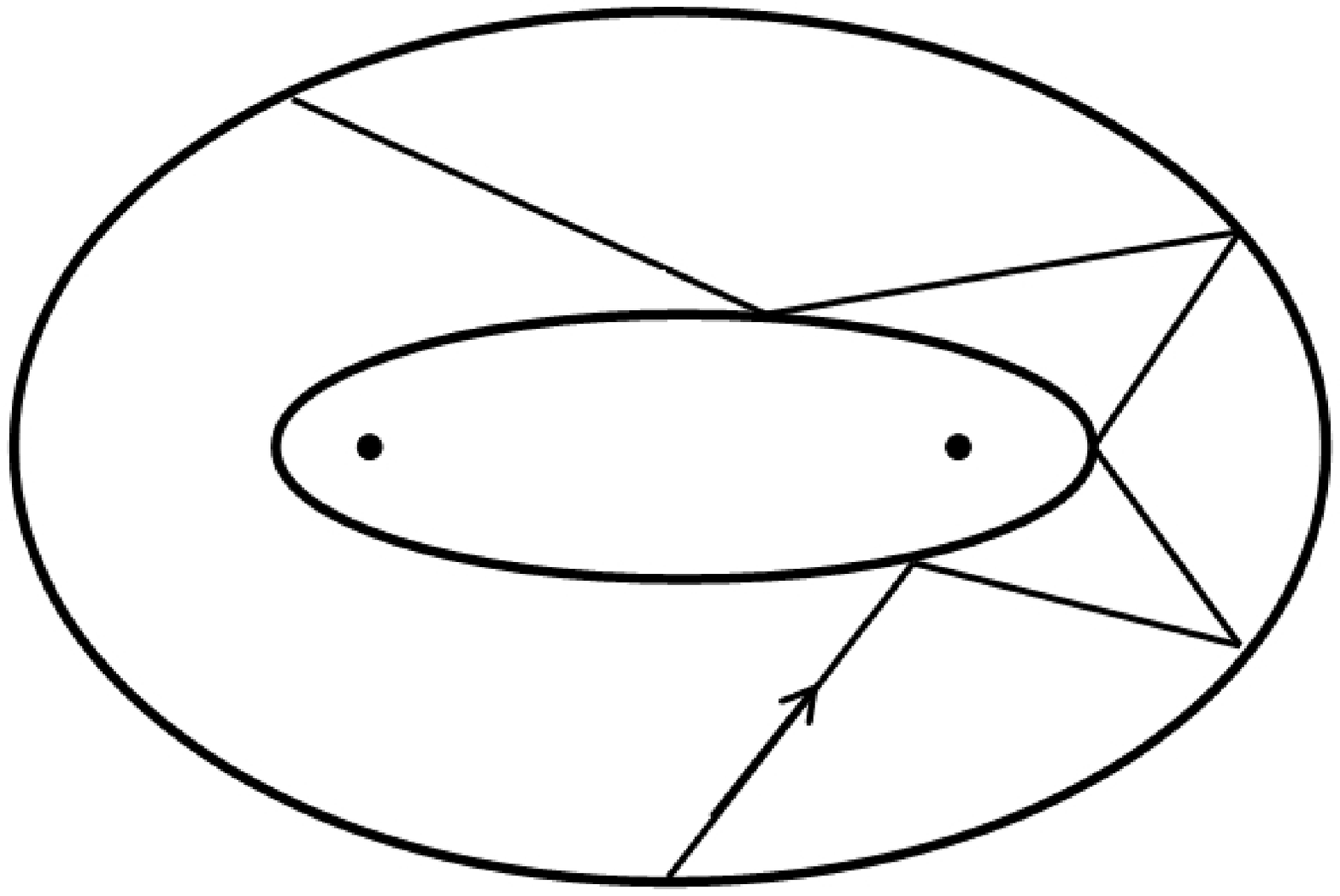}
\hskip1cm
\includegraphics[width=2.7cm,height=4cm]{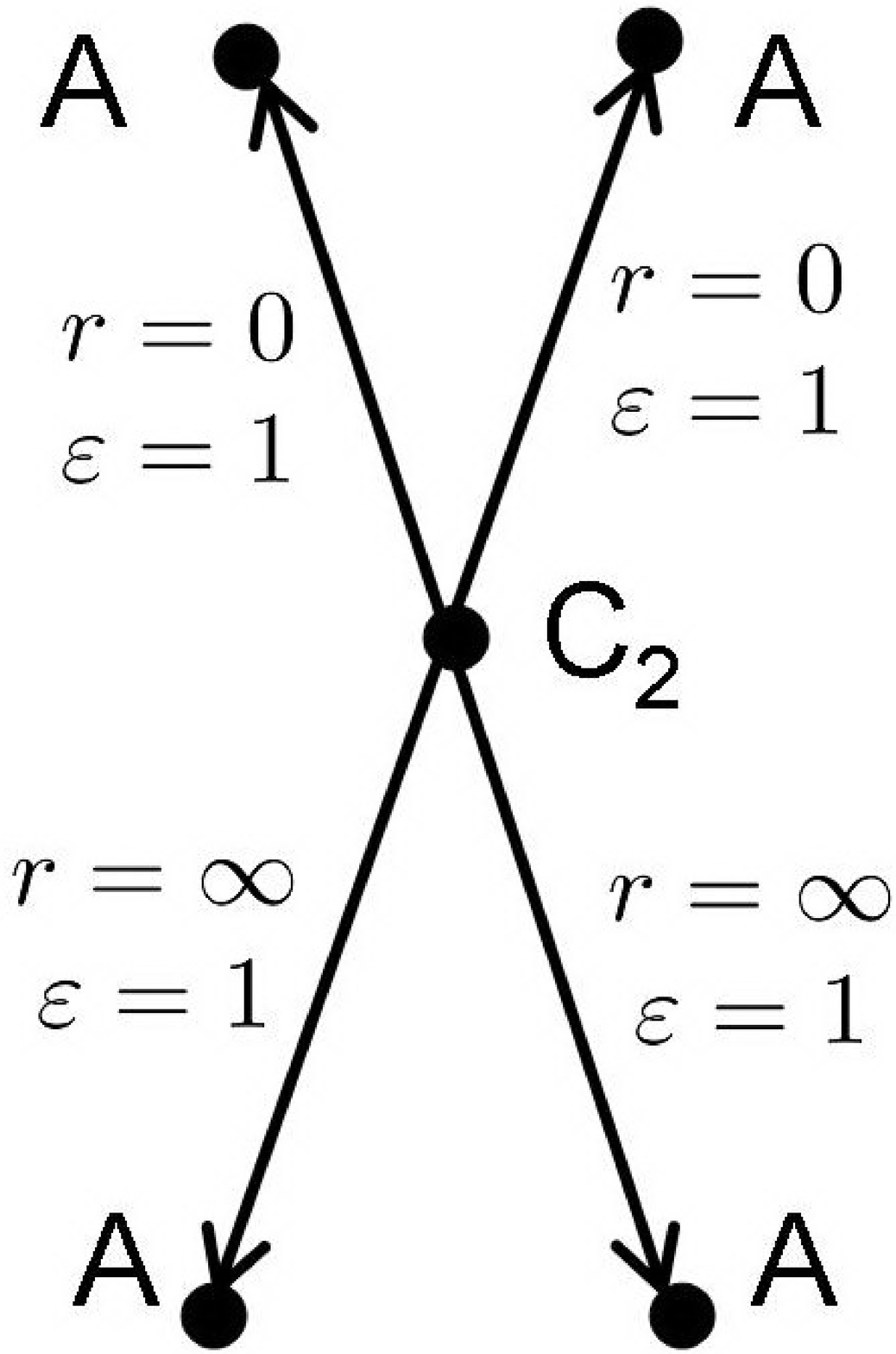}
{\caption{Billiard between two confocal ellipses and its Fomenko
graph}\label{fig:fom2}}
\end{figure}
\end{proposition}

\begin{proof}
As in Proposition \ref{prop:bilijar1}, the lower $\mathbf A$-atoms
correspond to the limit flows on the boundary of the outer
ellipse.

\smallskip

The level set for the caustic $\mu=b$ contains two periodic
trajectories placed on the $x$-axis and all trajectories such that
the continuations of their segments contain the foci. Let us note
that such a trajectory is heteroclinic and placed only in one of
the half-planes with the $x$-axis as the edge. Thus, this level
set has four separatrices: two in the upper half-plane, two in the
lower one. In each half-plane, one separatrix contains
trajectories tending to the left periodic trajectory on the
$x$-axis when time tends to $\infty$ and to the right one in
$-\infty$, and opposite for orbits on the other separatrix.
Liouville tori corresponding to elliptic caustics tend when $\mu$
tends to $b$, to two separatrices with the same movement direction
-- clockwise or counterclockwise.

\smallskip

For hyperbolic caustics, the billiard table become split into two
domains, symmetric with respect to the $x$-axis. Each family of
Liouville tori tend to the two separatrices from the level $\mu=b$
that are contained in the same half-plane.

\smallskip

This analysis shows that the atom that describes level set $\mu=b$
is $\mathbf C_2$.

\smallskip

The upper $\mathbf A$-atoms correspond to the periodic orbits
along $y$-axis.
\end{proof}

Now we will consider the domain is determined by an ellipse and a
confocal hyperbola. Let us note that the phase space in the cases,
where the boundary is not smooth at every point, must be
considered with a special attention. Since confocal conics are
always orthogonal in the intersection points, the periodic flows
along smooth arcs may appear as limits of the billiard motion,
when the caustic tend to the conic containing the arc. Thus,
although the tangent space to the boundary in the intersection
points is not defined, we will take that the ``allowed'' velocity
vectors in such points are those tangent to the curves containing
these points, with the opposite vectors identified with each
other.

\begin{proposition}\label{prop:ellipse.hyp}
Consider the billiard domain with the border composed of an
ellipse and a confocal hyperbola. Then:
\begin{itemize}

\item[1] If the domain is outside hyperbola, then the isoenergy
manifold is represented by the graph on Figure \ref{fig:bilijar3}.
\begin{figure}[h]
\centering
\includegraphics[width=5cm,height=4cm]{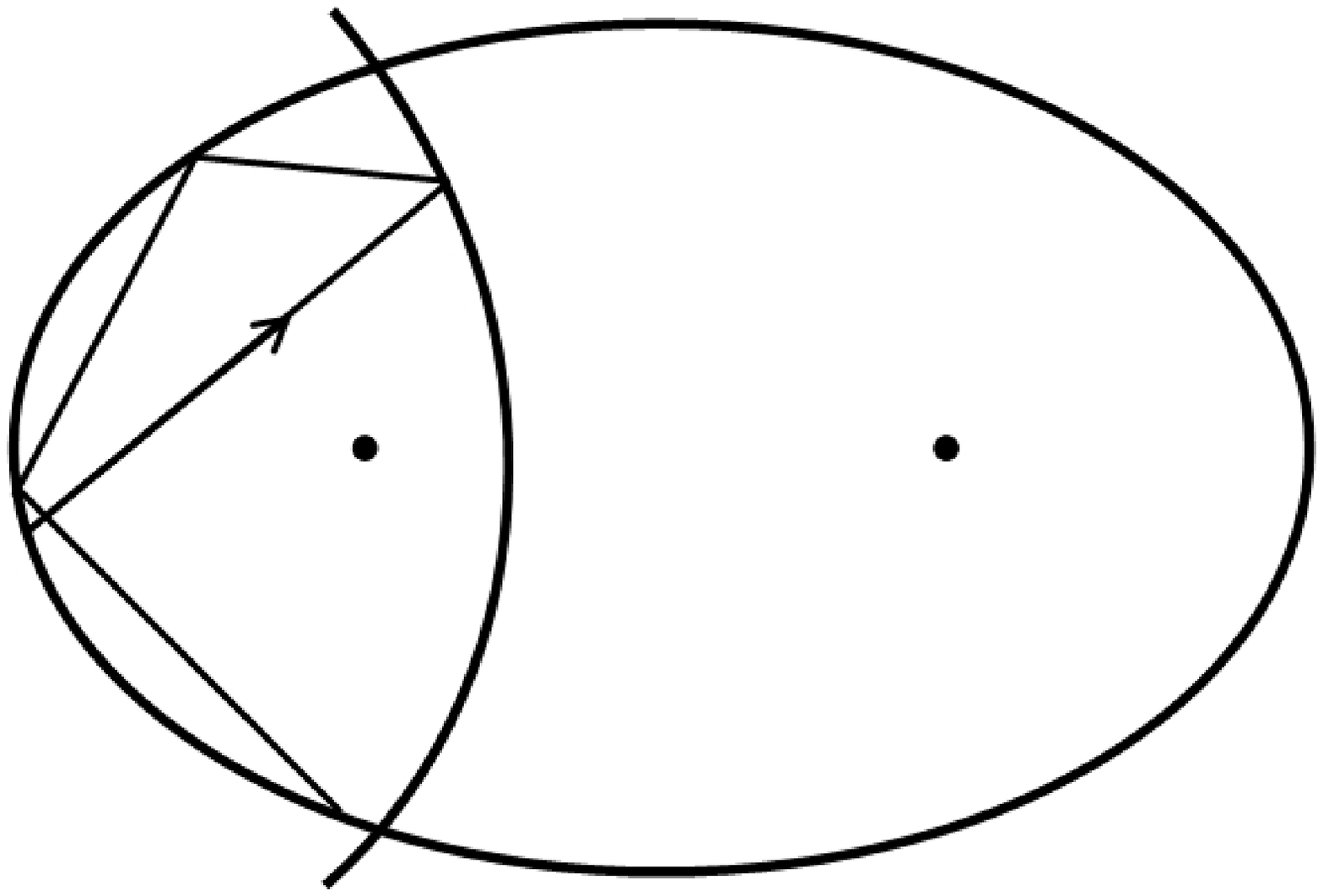}
\hskip1cm
\includegraphics[width=2.5cm,height=4cm]{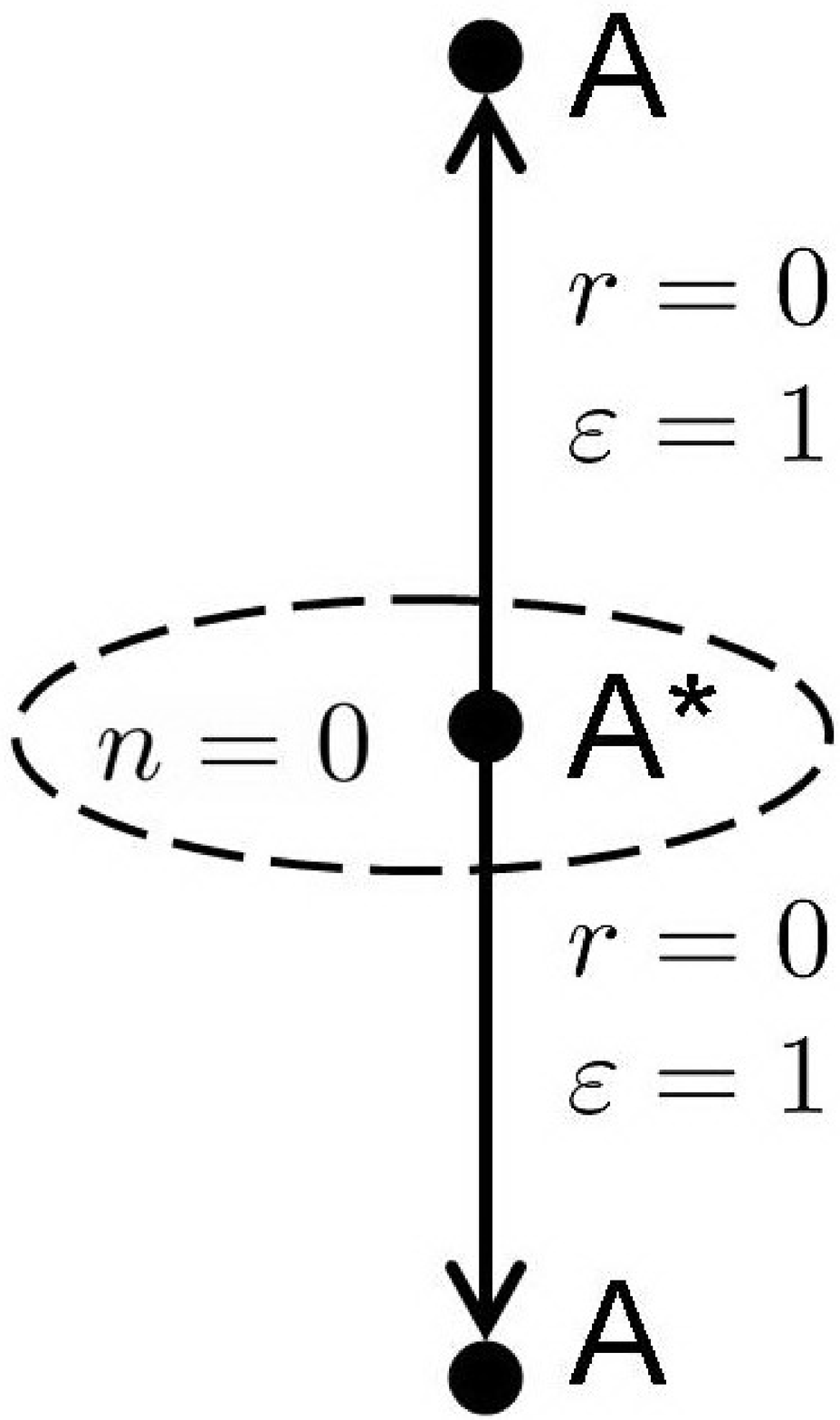}
{\caption{Billiard between ellipse and one branch of hyperbola and
its Fomenko graph}\label{fig:bilijar3}}
\end{figure}

\item[2] If the domain is inside hyperbola, then the isoenergy
manifold is represented by the graph on Figure \ref{fig:fom4}.
\begin{figure}[h]
\centering
\includegraphics[width=5cm,height=4cm]{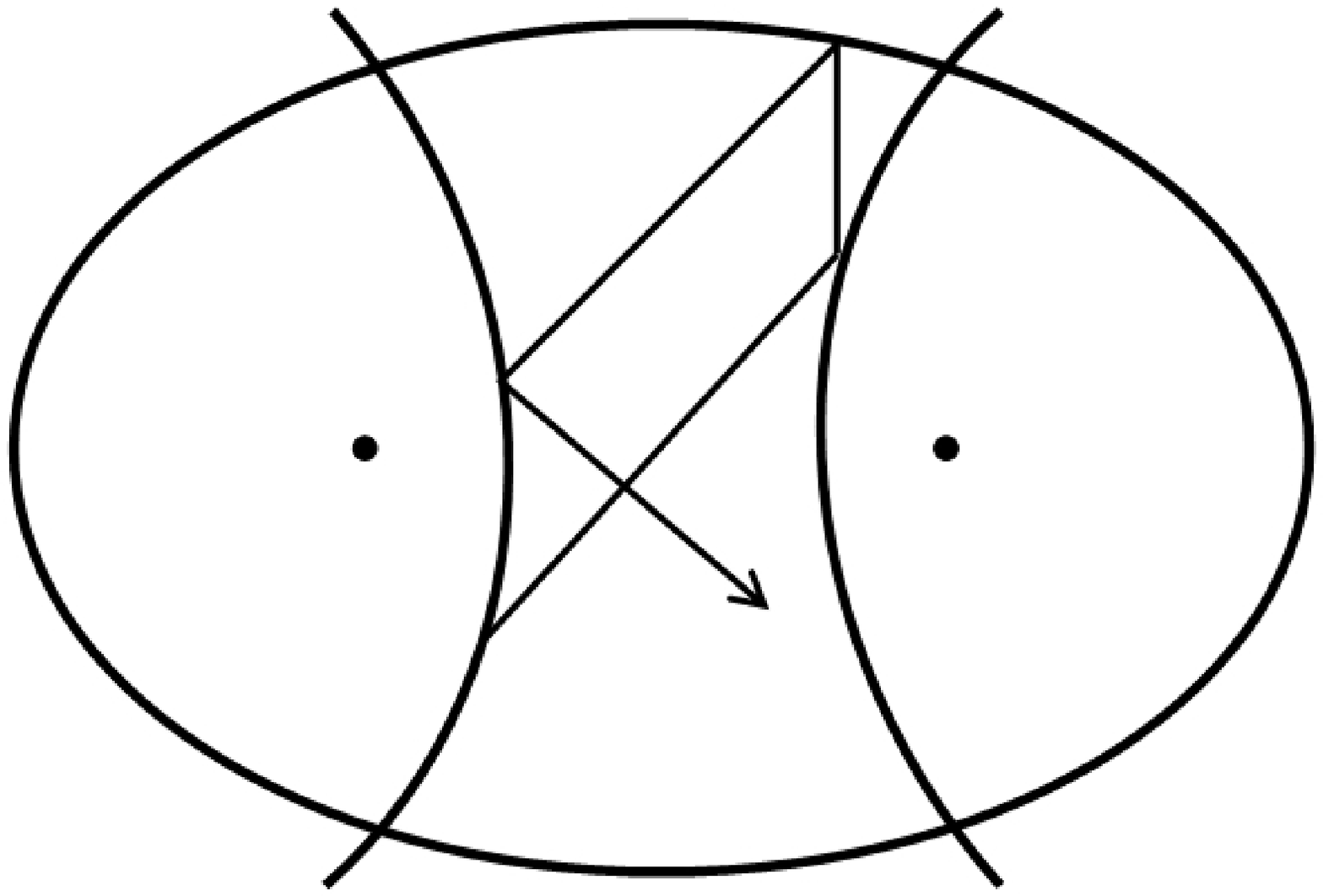}
\hskip1cm
\includegraphics[width=3cm,height=4cm]{fom1.ps}
{\caption{Billiard inside ellipse and hyperbola and its Fomenko
graph}\label{fig:fom4}}
\end{figure}
\end{itemize}
\end{proposition}

\begin{proof}
In the first case, each of the $\mathbf A$-atoms corresponds to
the limit periodic orbit along one of the smooth arcs constituting
the boundary of the billiard domain. The $\mathbf A^*$-atom
represents the periodic orbit along $x$-axis and its homoclinic
trajectories.

\smallskip

In the second case, the lower $\mathbf A$-atoms correspond to the
limit motion on two arcs of the ellipse, the $\mathbf B$-atom to
the periodic orbit on $x$-axis and its homoclinic orbits, and the
upper $\mathbf A$ atom to periodic orbit along $y$-axis. It is
interesting to note that this system is Liouville equivalent to
the billiard within the ellipse (compare Figures \ref{fig:fom1}
and \ref{fig:fom4}).
\end{proof}

We conclude this section by analyzing an example when the border
of the billiard table is continuously changed, to see how the
bifurcations in the isoenergy manifolds appear.

\begin{proposition}\label{prop:eeh}
Consider a billiard domain limited by two confocal ellipses and
two arcs contained in different branches of a confocal hyperbola.

\begin{itemize}

\item[1] If the domain is inside the hyperbola, then the isoenergy
manifold is represented by the graph on Figure \ref{fig:fom5}.
\begin{figure}[h]
\centering
\includegraphics[width=5cm,height=4cm]{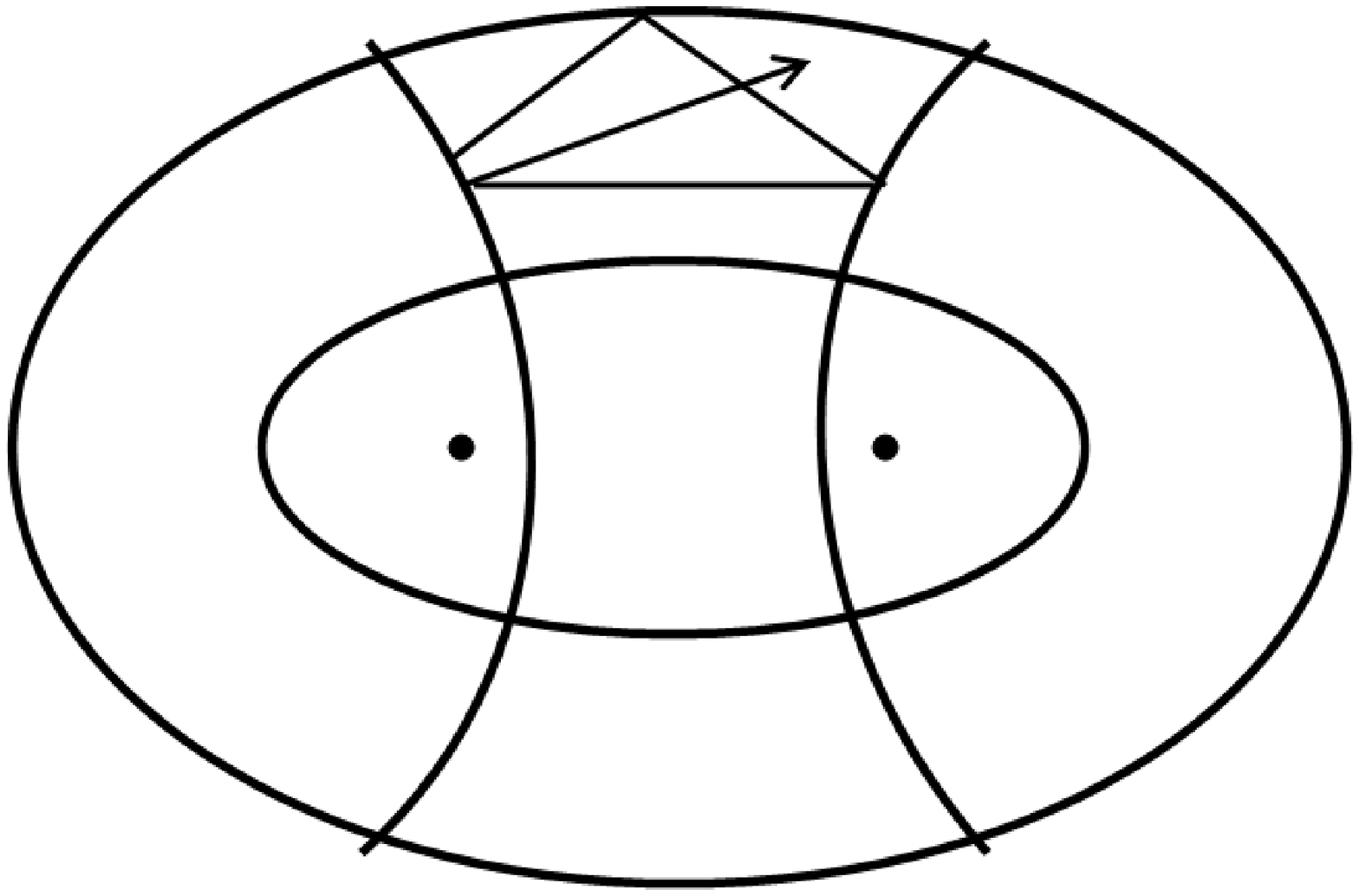}
\hskip1cm
\includegraphics[width=2.8cm,height=4cm]{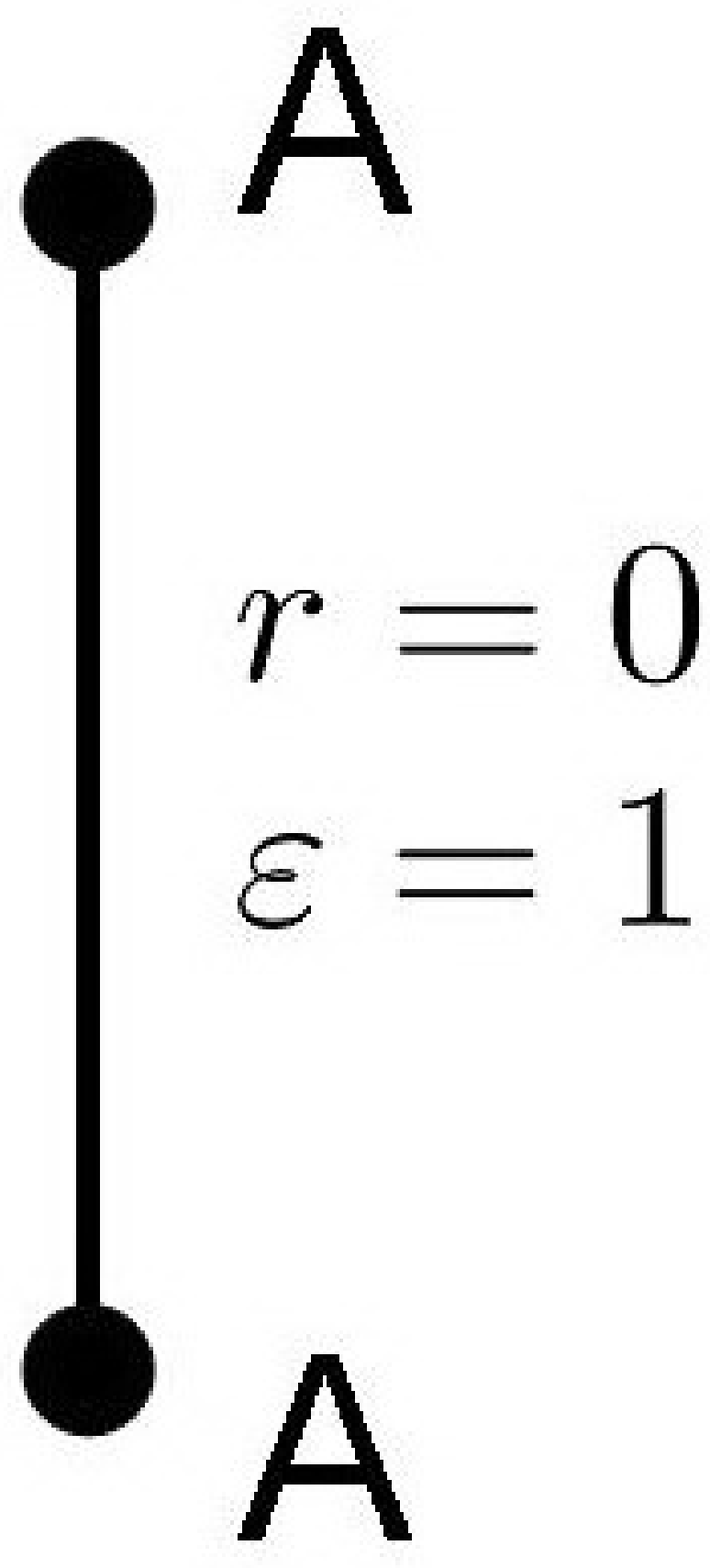}
{\caption{Billiard motion between two ellipses and hyperbola and
the corresponding Fomenko graph}\label{fig:fom5}}
\end{figure}

\item[2] If the $x$-axis is a part of the boundary, then the
isoenergy manifold is represented by the graph on Figure
\ref{fig:fom6}. Note that $\mathbf V$ corresponds to a degenerated
singular leaf containing two orientable periodic orbits and two
heteroclinic separatrices.
\begin{figure}[h]
\centering
\includegraphics[width=5cm,height=4cm]{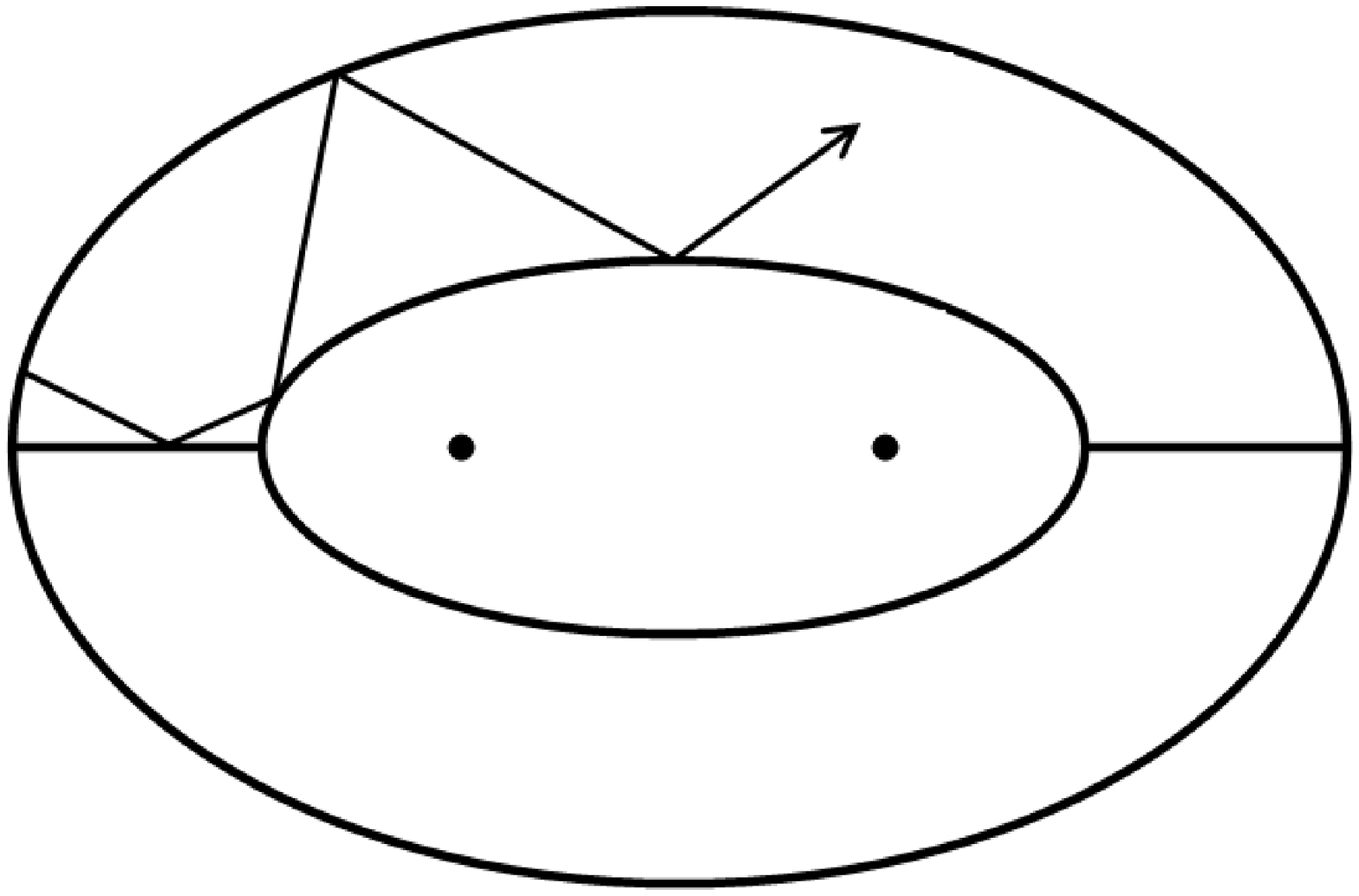}
\hskip1cm
\includegraphics[width=2cm,height=4cm]{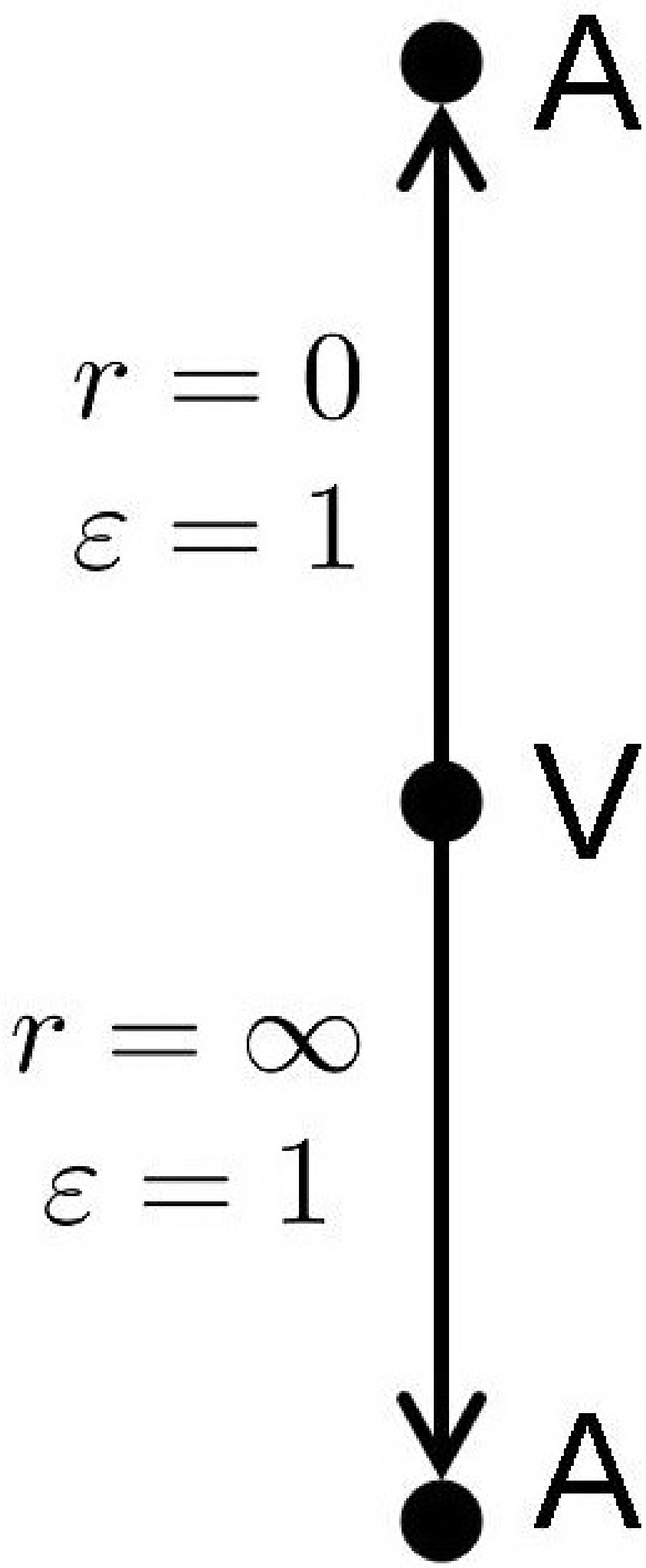}
{\caption{Billiard motion between two ellipses and a degenerate
hyperbola with the Fomenko graph}\label{fig:fom6}}
\end{figure}

\item[3] If the domain is as shown on the left side of Figure
\ref{fig:fom7}, then the isoenergy manifold is represented by the
graph on the right side of Figure \ref{fig:fom7}.
\begin{figure}[h]
\centering
\includegraphics[width=5cm,height=4cm]{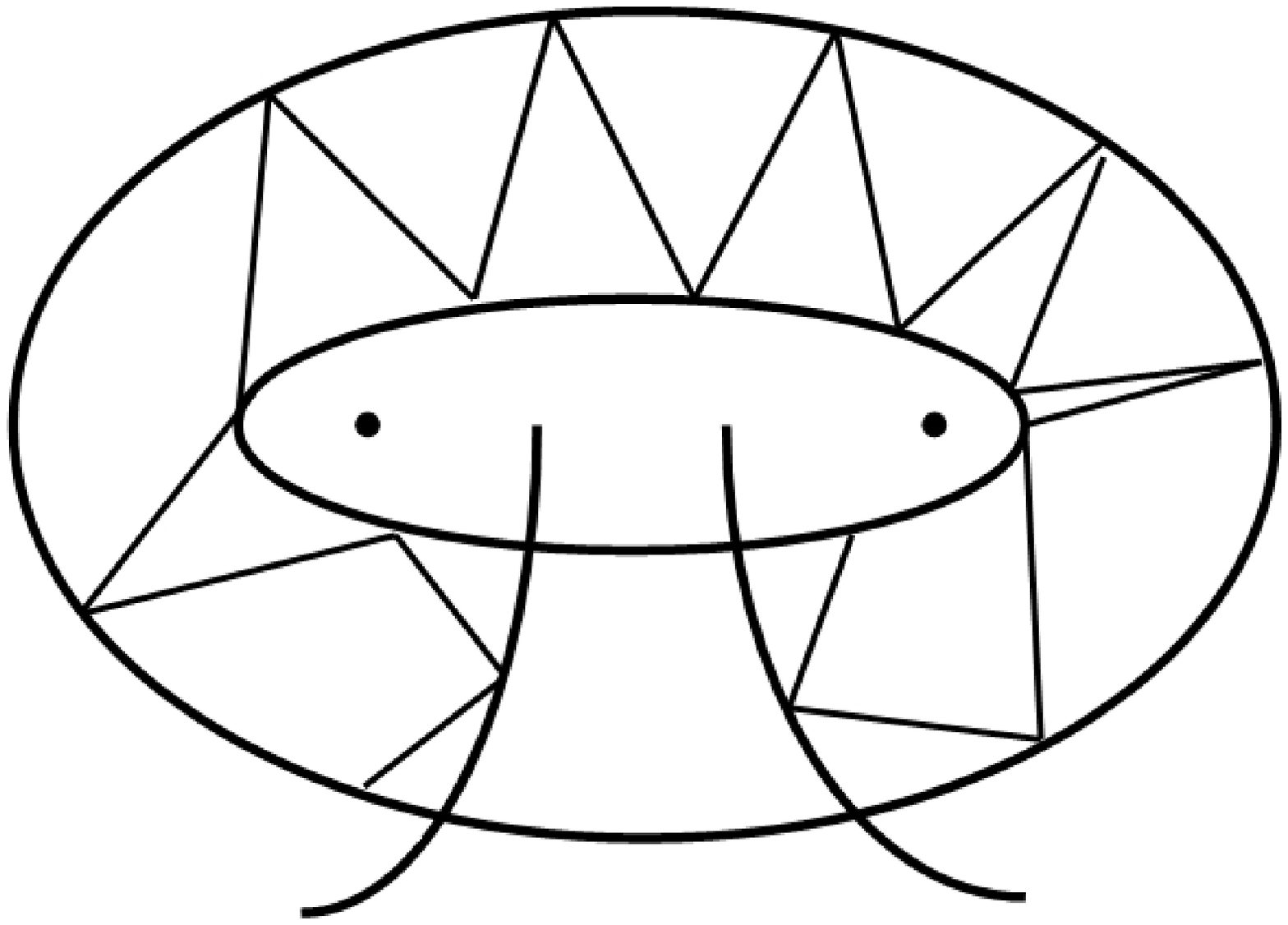}
\hskip1cm
\includegraphics[width=3.5cm,height=4cm]{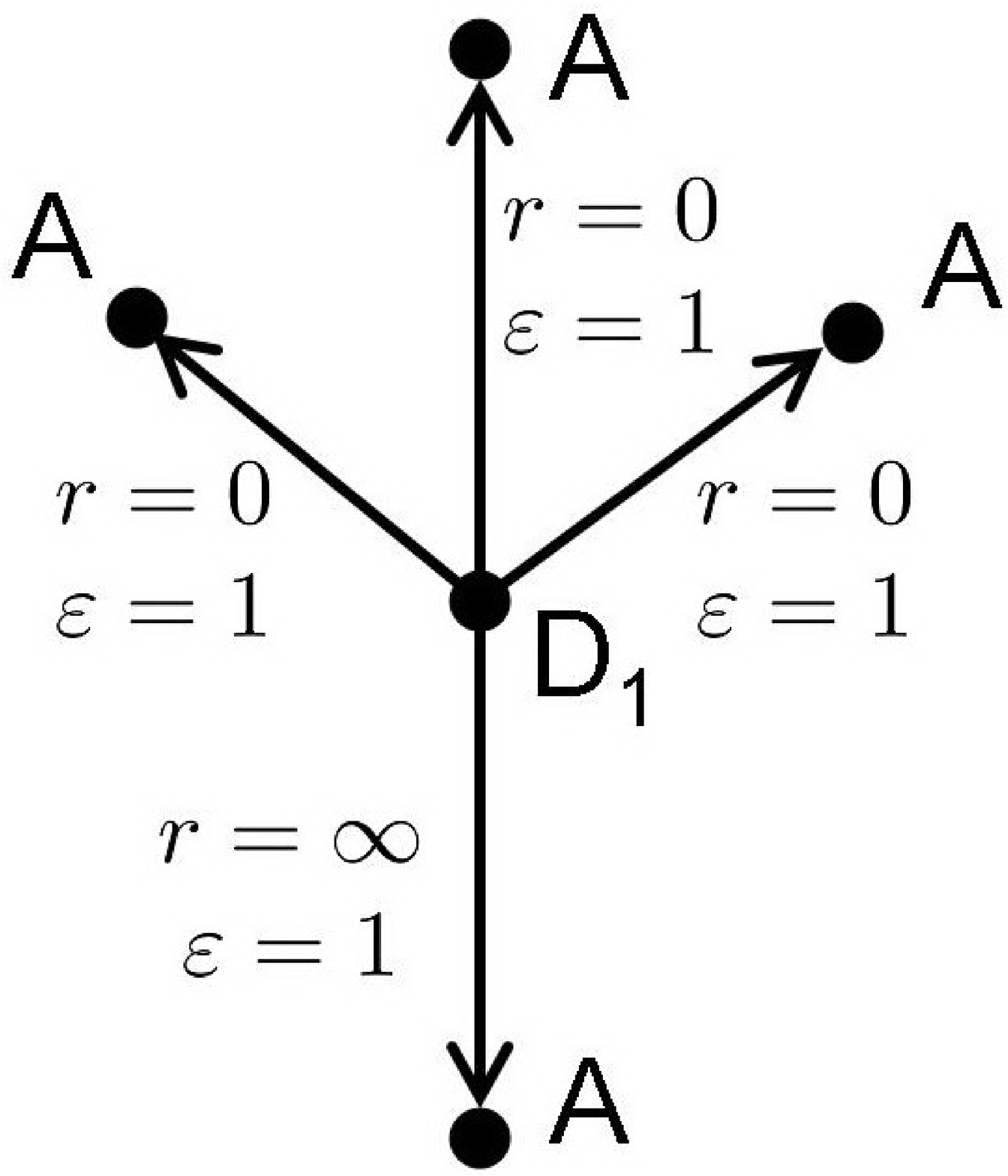}
{\caption{Billiard motion between two ellipses and hyperbola and
the corresponding Fomenko graph}\label{fig:fom7}}
\end{figure}
\end{itemize}
\end{proposition}

\begin{remark}
Near-integrable bifurcations of billiards from Propositions
\ref{prop:ellipse.hyp} and \ref{prop:eeh} can be studied by tools
developed in \cite{TuraevRomKedar2003}.
\end{remark}

\section{Billiards on Ellipsoids and Liouville surfaces}\label{sec:na.elipsoidu}

In this section, we are going to analyze the topology of the
billiard motion on the ellipsoid in $\mathbf E^3$, with the
boundary cut by a confocal quadric surface. Next, we will consider
the billiards within generalized ellipses on Liouville surfaces.
Obtained results are going to be compared with those from the
previous section.

\subsection{Topology of Geodesic Motion on Ellipsoid in $\mathbf
E^3$}

Since the segments of billiard trajectories on the ellipsoid are
placed on geodesic lines, it is essential to consider topology of
the isoenergy surfaces for the geodesic motion on the ellipsoid.
This topology is completely described in \cite{BolFom1994}.

\smallskip

We suppose that the ellipsoid in $\mathbf E^3$ is given by the
equation:
\begin{equation}\label{eq:elipsoid}
\mathcal E\ :\ \frac{x^2}a+\frac{y^2}b+\frac{z^2}c=1, \qquad
0<c<b<a.
\end{equation}

\begin{theorem}\label{th:jacobi}[\cite{BolFom1994, BolFomBOOK}]
The Fomenko graph for the Jacobi problem of geodesic lines on an
ellipsoid is presented in Figure \ref{fig:jakobi}.
\begin{figure}[h]
\centering
\includegraphics[width=3cm,height=4cm]{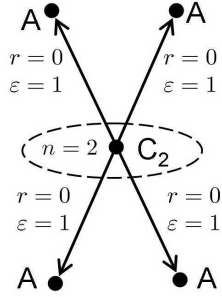}
{\caption{Fomenko graph for the Jacobi problem}\label{fig:jakobi}}
\end{figure}
The rotation function that correspond to the lower and upper edges
of the graph are:
$$
\rho_{\text{\rm lower}}(\alpha)
 =
 \frac
{\int_c^{\alpha}\Phi(\lambda,\alpha)d\lambda}%
{\int_b^a\Phi(\lambda,\alpha)d\lambda}\ \ (\alpha\in(c,b)),
 \quad
\rho_{\text{\rm upper}}(\alpha)
 =
 \frac
{\int_c^b\Phi(\lambda,\alpha)d\lambda}%
{\int_{\alpha}^a\Phi(\lambda,\alpha)d\lambda}\ \ (\alpha\in(b,a)),
$$
where
$$
\Phi(\lambda,\alpha)
 =
\frac{\lambda}{\sqrt{-\lambda(a-\lambda)(b-\lambda)(c-\lambda)(\alpha-\lambda)}}.
$$
\end{theorem}

It is interesting to remark that this theorem is used to prove that
the Jacobi problem of geodesic lines on an ellipsoid and the Euler
case of the rigid body motion are orbitally equivalent (see
\cite{BolFom1994,BolFomBOOK}).

\subsection{Billiards on Ellipsoid in $\mathbf E^3$}

A generalized ellipse on the surface of the ellipsoid
(\ref{eq:elipsoid}) is the intersection of this ellipsoid with a
confocal hyperboloid. Let us note that such an intersection
consists of two disjoint closed curves, which are curvature lines
on the ellipsoid. This intersection divides the surface of the
ellipsoid into three domains: two of them are simply connected and
congruent to each other, while the third one, placed between them
is $1$-connected.

\smallskip

\begin{figure}[h]
\centering
\includegraphics[width=6cm,height=5cm]{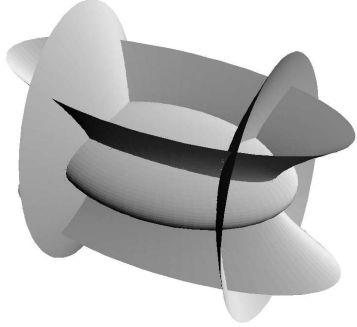}
{\caption{Confocal quadrics}\label{fig:konfokalna.familija}}
\end{figure}

More precisely, any surface confocal with the ellipsoid
(\ref{eq:elipsoid}) is given by the equation of the form (see
Figure \ref{fig:konfokalna.familija}):
\begin{equation}\label{eq:konfokalna.familija}
\mathcal Q_{\lambda}\ :\
Q_{\lambda}=\frac{x^2}{a-\lambda}+\frac{y^2}{b-\lambda}+\frac{z^2}{c-\lambda}=1.
\end{equation}

Let the Jacobi coordinates $(\lambda_1,\lambda_2,\lambda_3)$
related to this system of confocal quadric surfaces be ordered by
the condition $\lambda_1>\lambda_2>\lambda_3$.

\smallskip

\begin{figure}[h]
\centering
\includegraphics[width=8.3cm,height=4.4cm]{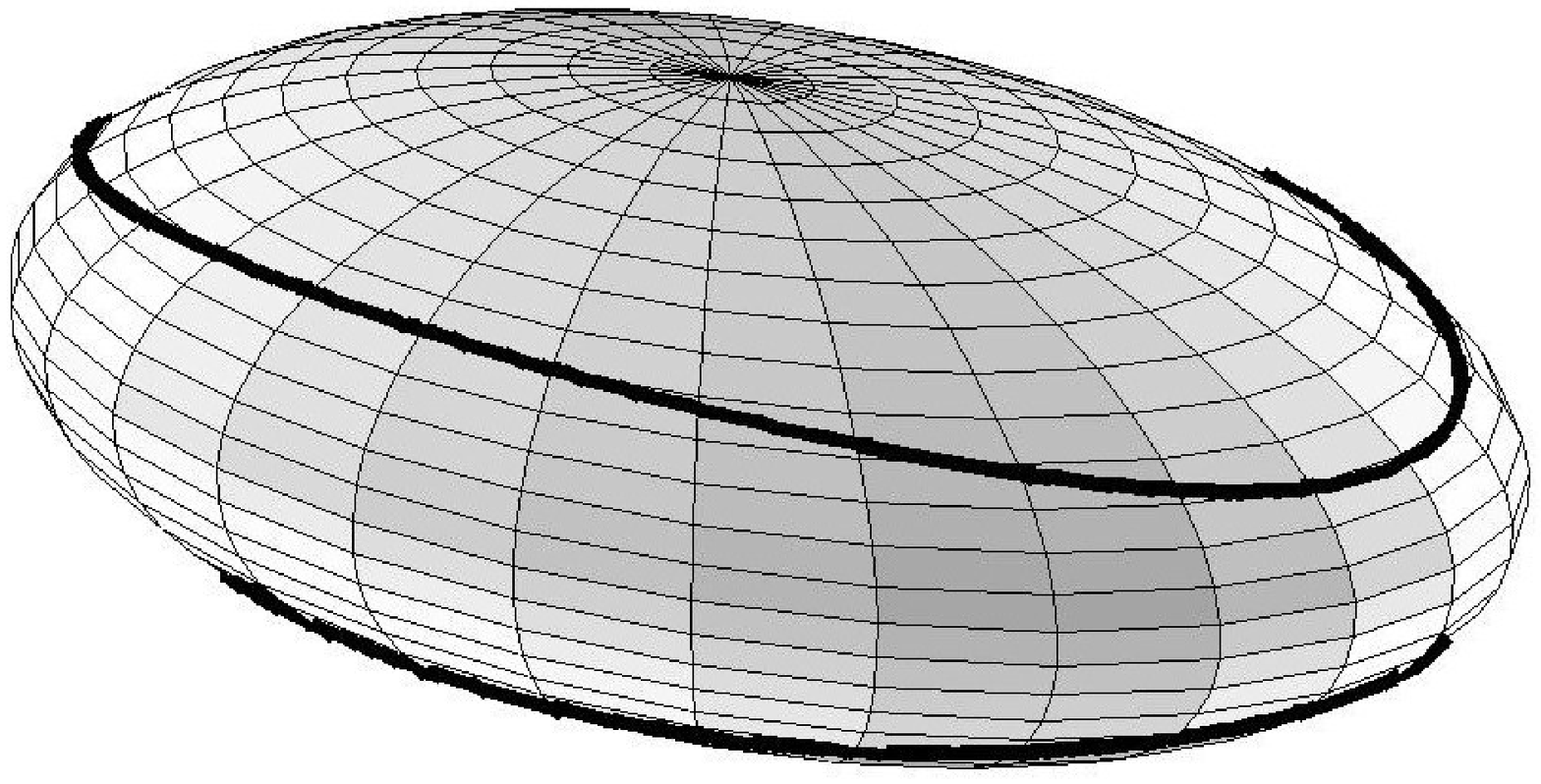}
{\caption{Intersection of ellipsoid with one-sheeted
hyperboloid}\label{fig:presek1}}
\end{figure}

Then the one-sheeted hyperboloid $\mathcal Q_{\beta}$,
$(c<\beta<b)$, cuts out the following three domains on the surface
of $\mathcal E$: $\Omega_1^{\beta}=\{\lambda_2>\beta,\ z>0\}$,
$\Omega_2^{\beta}=\{\lambda_2>\beta,\ z<0\}$ and
$\Omega_3^{\beta}=\{\lambda_2<\beta\}$. The first two domains are
symmetric with respect to the $xy$-plane, and the third one is the
ring placed between them on $\mathcal E$, as shown on Figure
\ref{fig:presek1}.

\smallskip

\begin{figure}[h]
\centering
\includegraphics[width=8.3cm,height=4.4cm]{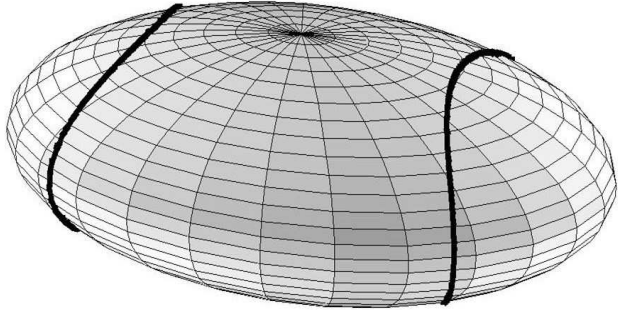}
{\caption{Intersection of ellipsoid with two-sheeted
hyperboloid}\label{fig:presek2}}
\end{figure}

Similarly, the two-sheeted hyperboloid $\mathcal Q_{\gamma}$,
$(b<\gamma<a)$, determines the domains
$\Omega_1^{\gamma}=\{\lambda_1<\gamma,\ x>0\}$,
$\Omega_2^{\gamma}=\{\lambda_1<\gamma,\ x<0\}$ and
$\Omega_3^{\gamma}=\{\lambda_1>\gamma\}$, see Figure
\ref{fig:presek2}.

\begin{proposition}
The isoenergy manifolds corresponding to the billiard systems
within the domains $\Omega_1^{\beta}$ and $\Omega_1^{\gamma}$ on
$\mathcal E$ is represented by the Fomenko graph on Figure
\ref{fig:fom1}.

The rotation functions for the billiard within $\Omega_1^{\beta}$,
for the lower and upper edges of the Fomenko graph respectively,
are:
$$
 \rho_{\text{\rm lower}}^{\beta}(\alpha)=\frac
{\int_{\beta}^{\alpha}\Phi(\lambda,\alpha)}%
{\int_b^a\Phi(\lambda,\alpha)}\ \ (\alpha\in(\beta,b)),
 \quad
\rho_{\text{\rm upper}}^{\beta}(\alpha)=\frac
{\int_{\beta}^b\Phi(\lambda,\alpha)}%
{\int_{\alpha}^a\Phi(\lambda,\alpha)}\ \ (\alpha\in(b,a)).
$$

The rotation functions for the billiard within
$\Omega_1^{\gamma}$, for the lower and upper edges of the Fomenko
graph respectively, are:
$$
 \rho_{\text{\rm lower}}^{\gamma}(\alpha)=\frac
{\int_{\alpha}^{\gamma}\Phi(\lambda,\alpha)}%
{\int_c^b\Phi(\lambda,\alpha)}\ \ (\alpha\in(b,\gamma)),
 \quad
 \rho_{\text{\rm upper}}^{\gamma}(\alpha)=\frac
{\int_b^{\gamma}\Phi(\lambda,\alpha)}%
{\int_c^{\alpha}\Phi(\lambda,\alpha)}\ \ (\alpha\in(c,b)).
$$
$\Phi$ is defined as in Theorem \ref{th:jacobi}.\end{proposition}

\subsection{Billiard within an Ellipse on a Liouville surface}

Now, let us state the main result of this section.

\begin{theorem}
All billiard systems within ellipse on an arbitrary Liouville
surface are Liouville equivalent. The Fomenko graph corresponding
to an isoenergy manifold of such a system is represented on Figure
\ref{fig:fom1}.
\end{theorem}

\begin{proof}
Families of confocal ellipses and hyperbolas, together with
corresponding billiard systems, were defined and described by
Darboux \cite{DarbouxSUR} (for a recent account on this topic, see
\cite{DragRadn2006,DragRadn2008}). Since such billiard have all
topological properties analogous to billiards within an ellipse in
the Euclidean plane, the theorem is proved.
\end{proof}

\section{Billiards inside Ellipsoid in $\E^3$}\label{sec:vise.dimenzija}

Here, we are going to analyze the topology of the billiard motion
within an ellipsoid in $\mathbf E^3$. We suppose that the equation
of the ellipsoid is (\ref{eq:elipsoid}).

\smallskip

\begin{figure}[h]
\centering
\includegraphics[width=5cm,height=4cm]{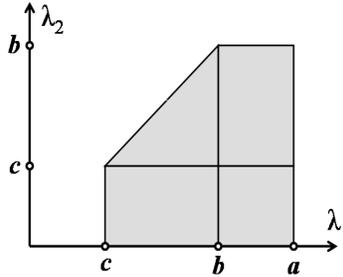}
{\caption{The bifurcation set for the billiard within an ellipsoid
in $\mathbf E^3$}\label{fig:bife3}}
\end{figure}

Each trajectory of the billiard motion within $\mathcal E$ has
exactly two caustics from the confocal family $\mathcal
Q_{\lambda}$. Suppose these two surfaces are $\mathcal
Q_{\lambda_1}$ and $\mathcal Q_{\lambda_2}$,
$\lambda_2\le\lambda_1$. Then the bifurcation set of an isoenergy
surface of the system is given on Figure \ref{fig:bife3}.

\smallskip

The bifurcations of the Liouville tori for this system were
investigated in \cite{Knorr1985, Audin1994, DelshamsFedRR2001}.

\smallskip

Now, let us explain in detail how the behaviour of the billiard
motion is connected to the diagram on Figure \ref{fig:bife3}.

\smallskip

Over each point in the gray area on the diagram, there is one or
two $3$-dimensional Liouville tori. The marked edges will
correspond to the degenerated level sets.

\smallskip

Now, let us describe in detail the level sets lying over different
points in the bifurcation set on the Figure \ref{fig:bife3}.

\subsection{Regular Leaves}

Over each point inside rectangles $[c,b]\times[0,c]$,
$[b,a]\times[0,c]$ and the triangle with vertices $(c,c)$,
$(b,c)$, $(b,b)$, there are two Liouville tori $\mathbf T^3$. Over
each point inside rectangle $[b,a]\times[c,b]$, there is only one
$\mathbf T^3$.

\begin{itemize}
\item
Points inside $[c,b]\times[0,c]$ correspond to the motion
with an ellipsoid and a one-sheeted hyperboloid as caustics. One
Liouville torus is formed by the trajectories winding in the
positive direction around the $z$-axis, and the other one by those
winding in the negative direction.

\item
Points inside $[b,a]\times[0,c]$ correspond to the case when one
caustic is ellipsoid and the other a two-sheeted hyperboloid. Each
Liouville torus is formed by the trajectories winding in one
direction around the $x$-axis.

\item
Points inside triangle $(c,c)-(b,c)-(b,b)$ correspond to the
motion with both caustics being different one-sheeted
hyperboloids. Each of the Liouville tori is formed by the
trajectories winding in one direction around the $x$-axis.

\item
Points inside rectangle $[b,a]\times[c,b]$ correspond to the case
when both caustics are hyperboloids, but of different type. Then,
all corresponding billiard trajectories form one Liouville torus.
\end{itemize}

\subsection{Singular Leaves}

Singular leaves are lying over edges of the diagram represented on
Figure \ref{fig:bife3}. They appear when one or both caustics are
degenerated.

\smallskip

First, let us clarify the notion of a degenerated quadric from
confocal family (\ref{eq:konfokalna.familija}), as well as the
geometrical meaning of the billiard motion with such a caustic.

\smallskip

\begin{figure}[h]
 \centering
\includegraphics[width=6cm,height=5cm]{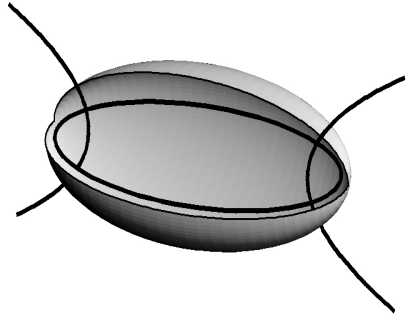}
{\caption{Ellipsoid $\mathcal E$ and degenerated quadrics
$\mathcal Q_b$, $\mathcal Q_c$}\label{fig:degenerisane}}
\end{figure}

The degenerated quadrics $\mathcal Q_{\lambda}$,
$\lambda\in\{a,b,c\}$ are the following curve lying in the
coordinate planes:
$$
\aligned
 &\mathcal Q_a\ :\ -\frac{y^2}{a-b}-\frac{z^2}{a-c}=1,\quad x=0,\\
 &\mathcal Q_b\ :\ \frac{x^2}{a-b}-\frac{z^2}{b-c}=1,\quad y=0,\\
 &\mathcal Q_c\ :\ \frac{x^2}{a-c}+\frac{y^2}{b-c}=1,\quad z=0.
\endaligned
$$
Notice that $\mathcal Q_c$ is an ellipse in the $xy$-plane,
$\mathcal Q_b$ is a hyperbola in the $xz$-plane, while $\mathcal
Q_c$ is, in the real case, the empty set. Nevertheless, we will
consider $\mathcal Q_a$ as an abstract curve in the $yz$-plane.
The degenerated quadrics are depicted on Figure
\ref{fig:degenerisane}.

\smallskip

The billiard motion with some of these three degenerated caustics
$\mathcal Q_{\lambda}$ is either such that each segment of the
trajectory intersects the curve $\mathcal Q_{\lambda}$ or the
whole trajectory is lying in the coordinate plane containing this
curve. In the latter case, the motion reduces to the plane
billiard one of the ellipses:
$$
\aligned
 &\mathcal E_a\ :\ \frac{y^2}b+\frac{z^2}c=1,\quad x=0,\\
 &\mathcal E_b\ :\ \frac{x^2}a+\frac{z^2}c=1,\quad y=0,\\
 &\mathcal E_c\ :\ \frac{x^2}a+\frac{y^2}b=1,\quad z=0.
\endaligned
$$
Notice that, in the case $\lambda=a$, only plane trajectories
exist.

\smallskip

\begin{figure}[h]
\centering
\includegraphics[width=6cm,height=4cm]{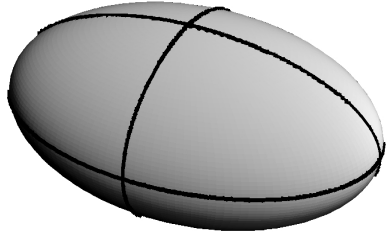}
{\caption{Ellipsoid $\mathcal E$ and ellipses $\mathcal E_a$,
$\mathcal E_b$, $\mathcal E_c$}\label{fig:koordinatne}}
\end{figure}

Each of these three ellipses is the intersection of $\mathcal E$
with one of the coordinate hyper-planes (see Figure
\ref{fig:koordinatne}).

\smallskip

Besides these cases with a degenerated caustic, there one more
special case -- when one of the caustics is the boundary ellipsoid
$\mathcal E$. This case can be considered as the limit case with
the caustic $\mathcal Q_{\lambda}$, $\lambda\to0_+$, and its
trajectories are geodesic lines on $\mathcal E$.

\smallskip

Now, let us analyze the edges of the bifurcation diagram on Figure
\ref{fig:bife3}.

\subsubsection*{Edge $[c,a]\times\{0\}$}
As already mentioned, the motion corresponding to this edge is the
geodesic motion on $\mathcal E$. Thus, the topology of the
corresponding subset in the isoenergy manifold is completely
described by the Fomenko graph on Figure \ref{fig:jakobi}.

\smallskip

Let us summarize this case:
\begin{itemize}
\item
Point $(c,0)$ corresponds to the motion along the curve $\mathcal
E_c$. There are two $1$-dimensional tori $\mathbf T^1$ lying over
the point $(c,0)$ of the bifurcation diagram -- each one
corresponding to the flow in one direction.

\item
Inner points of the segment $[c,b]$ on $\lambda_1$-axis correspond
to the geodesic motion on $\mathcal E$ with a one-sheeted
hyperboloid as caustic. Such geodesic lines fill the ring between
two intersection curves of $\mathcal E$ and the hyperboloid. Over
each point inside the segment, there are two tori $\mathbf T^2$ --
each one corresponding to winding in one direction around the
$z$-axis.

\item
The level set corresponding to $(b,0)$ contains two periodic
trajectories and four two-dimensional separatrices. The periodic
trajectories are placed along the curve $\mathcal E_b$.
Trajectories on the separatrices are passing through the umbilical
points of the ellipsoid.

\item
Inner points of the segment $[b,a]$ on $\lambda_1$-axis correspond
to the geodesic motion on $\mathcal E$ with a two-sheeted
hyperboloid as caustic. Such geodesic lines fill the ring between
two intersection curves of $\mathcal E$ and the hyperboloid. Over
each point inside the segment, there are two tori $\mathbf T^2$ --
each one corresponding to winding in one direction around the
$x$-axis.

\item
Point $(a,0)$ corresponds to the motion along $\mathcal E_a$.
There are two $1$-dimensional tori $\mathbf T^1$ lying over this
point -- each one corresponding to the motion in one direction.
\end{itemize}

\subsubsection*{Edge $\{a\}\times[0,b]$}
This edge corresponds to the billiard motion with the degenerated
caustic $\mathcal Q_a$. All trajectories of such a motion are
placed in the $yz$-plane, thus this edge in fact corresponds to
the billiard within an ellipse. The topology of the set lying over
this edge is completely described by the Fomenko atom on Figure
\ref{fig:fom1}.

\subsubsection*{Edge $[b,a]\times\{b\}$}
This edge corresponds to the billiard motion with one degenerate
caustic $\mathcal Q_b$, and the other caustic being a two-sheeted
hyperboloid.

\subsection*{Acknowledgements}
\addcontentsline{toc}{section}{Acknowledgements}

We thank A.~V.~Bolsinov and V.~Rom-Kedar for interesting
discussions. This work is partially supported by the Serbian
Ministry of Science and Technology, Project no.\ 144014:
\textit{Geometry and Topology of Manifolds and Integrable Dynamical
Systems}.

\section*{Appendix: Fomenko graphs}

In the appendix, we are going to give a concise description of the
representation of isoenergy surfaces and some of their topological
invariants by Fomenko graphs. For all details, see
\cite{BolFom1994,BO2006} and references therein.

\smallskip

Let $(\mathcal{M},\omega,H,K)$ be an integrable system given on a
four-dimensional manifold with a symplectic form $\omega$. Here
$H$, $K$ denote independent functions on $\mathcal{M}$, commuting
with respect to the symplectic structure on $\mathcal{M}$.

\smallskip

\emph{Level sets} are subsets of $\mathcal M$ given by equations
$H=\const$, $K=\const$. The \emph{Liouville foliation} of
$\mathcal{M}$ is its decomposition into connected components of the
level sets. A leave of the foliations is \emph{regular} if $dH$,
$dK$ are independent at each its point, otherwise it is called
\emph{singular}. Singular leaves satisfying some additional
conditions will be called \emph{non-degenerate}. For these
non-degeneracy conditions, see \cite{BolFom1994}.

\smallskip

We suppose that \emph{isoenergy surface} $H=\const$ are compact,
thus each foliation leaf will also be compact. Thus, by the
Arnol'd-Liouville theorem \cite{ArnoldMMM}, each regular leaf is
diffeomorphic to the $2$-dimensional torus $\mathbf{T}^{2}$ and the
the motion in its neighborhood is completely described by, for
example, the action-angle coordinates. We say that isoenergy
surfaces are \textit{Liouville equivalent} if they are topologically
conjugate and the homeomorphism preserves their Liouville foliations

\smallskip

The set of topological invariants that describe completely
isoenergy manifolds containing regular and non-degenerate leaves
not containing fixed points of , up to their Liouville
equivalence, consists of:

\begin{itemize}
\item[$\bullet$] \textit{The oriented graph} $G$, whose vertices
correspond to the singular connected components of the level sets
of $K$, and edges to one-parameter families of Liouville tori;

\item[$\bullet$] \textit{The collection of Fomenko atoms}, such
that each atom marks exactly one vertex of the graph $G$;

\item[$\bullet$] \textit{The collection of pairs of numbers}
$(r_{i},\varepsilon_{i})$, with
$r_{i}\in([0,1)\cap\mathbf{Q})\cup\{\infty\}$,
$\varepsilon_{i}\in\{-1,1\}$, $1\le i\le n$. Here, $n$ is the
number of edges of the graph $G$ and each pair
$(r_{i},\varepsilon_{i})$ marks an edge of the graph;

\item[$\bullet$] \textit{The collection of integers}
$n_{1},n_{2},\dots,n_{s}$. The numbers $n_{k}$ correspond to
certain connected components of the subgraph $G^{0}$ of $G$.
$G^{0}$ consists of all vertices of $G$ and the edges marked with
$r_{i}=\infty$. The connected components marked with integers
$n_{k}$ are those that do not contain a vertex corresponding to an
isolated critical circle (an $\mathbf{A}$ atom) on the manifold
$\mathcal{Q}$.
\end{itemize}

Let us clarify the meaning of these invariants.

\smallskip

First, we are going to describe the construction of the graph $G$
from the manifold $\mathcal{Q}$. Each singular leaf of the
Liouville foliation corresponds to exactly one vertex of the
graph. If we cut $\mathcal{Q}$ along such leaves, the manifold
will fall apart into connected sets, each one consisting of
one-parameter family of Liouville tori. Each of these families is
represented by an edge of the graph $G$. The vertex of $G$ which
corresponds to a singular leaf $\mathcal{L}$ is incident to the
edge corresponding to the family $\mathcal{T}$ of tori if and only
if $\partial\mathcal{T}\cap\mathcal{L}$ is non empty.

\smallskip

Note that $\partial\mathcal{T}$ has two connected components, each
corresponding to a singular leaf. If the two singular leaves
coincide, then the edge creates a loop connecting one vertex to
itself.

\smallskip

Now, when the graph is constructed, one need to add the
orientation to each edge. This may be done arbitrarily, but, once
determined, the orientation must stay fixed because the values of
numerical Fomenko invariants depend on it.

\smallskip

The Fomenko atom which corresponds to a singular leaf
$\mathcal{L}$ of a singular level set is determined by the
topological type of the set $\mathcal{L}_{\varepsilon}$. The set
$\mathcal{L}_{\varepsilon}\supset{\mathcal{L}}$ is the connected
component of
 $\{\ p\in\mathcal{Q}\ |\ c-\varepsilon<K(p)<c+\varepsilon\ \}$,
where $c=K(\mathcal{L})$, and $\varepsilon>0$ is such that $c$ is
the only critical value of the function $K$ on $\mathcal{Q}$ in
interval $(c-\varepsilon,c+\varepsilon)$.

\smallskip

Let
 $\mathcal{L}_{\varepsilon}^{+} = \{\ p\in\mathcal{L}_{\varepsilon}\ |\ K(p)>c\ \}$,
 $\mathcal{L}_{\varepsilon}^{-} = \{\ p\in\mathcal{L}_{\varepsilon}\ |\ K(p)<c\ \}$.
Each of the sets $\mathcal{L}_{\varepsilon }^{+} $,
$\mathcal{L}_{\varepsilon}^{-}$ is a union of several connected
components, each component being a one-parameter family of
Liouville tori. Each of these families corresponds to the
beginning of an edge of the graph $G$ incident to the vertex
corresponding to $\mathcal{L}$.

\smallskip

Let us say a few words on the topological structure of the set
$\mathcal{L}$. This set consists of at least one fixed point or
closed one-dimensional orbit of the Poisson action $\Phi$ on
$\mathcal{M}$ and several (possibly none) two-dimensional orbits
of the action, which are called \emph{separatrices}.

\smallskip

The trajectories on each of these two-dimensional separatrices is
homoclinically or heteroclinically tending to the
lower-dimensional orbits. The Liouville tori of each of the
families in
 $\mathcal{L}_{\varepsilon}\setminus\mathcal{L}
=
 \mathcal{L}_{\varepsilon}^{+}\cup\mathcal{L}_{\varepsilon}^{-}$
tend, as the integral $K$ approaches $c$, to a closure of a subset
of the separatrix set.

\subsection*{Fomenko Atoms}

Fomenko and his school completely described and classified
non-degenerate leaves that do not contain fixed points of the
Hamiltonian $H$.

\smallskip

If $\mathcal{L}$ is not an isolated critical circle, then a
sufficiently small neighborhood of each $1$-dimensional orbit in
$\mathcal{L}$ is isomorphic to either two cylinders intersecting
along the base circle, and then the orbit is \textit{orientable},
or to two Moebius bands intersecting each other along the joint
base circle, then the $1$-dimensional orbit is
\textit{non-orientable}.

\smallskip

The number of closed one-dimensional orbits in $\mathcal{L}$ is
called \textit{the complexity} of the corresponding atom.

\subsection*{Fomenko atoms of complexity 1}
There are exactly three such atoms.

\subsubsection*{The atom $\mathbf{A}$.}

This atom corresponds to a normally elliptic singular circle,
which is isolated on the isoenergy surface $\mathcal{Q}$. A small
neighborhood of such a circle in $\mathcal{Q}$ is diffeomorphic to
a solid torus. One of the sets $\mathcal{L}_{\varepsilon}^{+}$,
$\mathcal{L}_{\varepsilon}^{-}$ is empty, the other one is
connected. Thus only one edge of the graph $G$ is incident with
the vertex marked with the letter atom $\mathbf{A}$.

\subsubsection*{The atom $\mathbf{B}$.}

In this case, $\mathcal{L}$ consists of one orientable normally
hyperbolic circle and two $2$-dimensional separatrices -- it is
diffeomorphic to a direct product of the circle $\mathbf{S}^{1}$
and the plane curve given by the equation $y^{2}=x^{2}-x^{4}$.
Because of its shape, we will refer to this curve as the
\textit{`figure eight'}. The set $\mathcal{L}_{\varepsilon
}\setminus\mathcal{L}$ has $3$ connected components, two of them
being placed in $\mathcal{L}_{\varepsilon}^{+}$ and one in
$\mathcal{L}_{\varepsilon}^{-}$, or vice versa. Let us fix that
two of them are in $\mathcal{L}_{\varepsilon }^{+}$. Each of these
two families of Liouville tori limits as $K$ approaches $c$ to
only one of the separatrices. The tori in
$\mathcal{L}_{\varepsilon }^{-}$ tend to the union of the
separatrices.

\subsubsection*{The atom $\mathbf{A}^{*}$.}

$\mathcal{L}$ consists of one non-orientable hyperbolic circle and
one $2$-di\-men\-sional separatrix. It is homeomorphic to the
smooth bundle over $\mathbf{S}^{1}$ with the `figure eight' as
fiber and the structural group consisting of the identity mapping
and the central symmetry of the `figure eight'. Both
$\mathcal{L}_{\varepsilon}^{+}$ and $\mathcal{L}_{\varepsilon
}^{-}$ are $1$-parameter families of Liouville tori, one limiting
to the separatrix from outside the `figure eight' and the other
from the interior part of the `figure eight'.

\subsection*{Fomenko atoms of complexity 2}
There are six such atoms. However, we give here the description
only of those appearing in the examples of this paper, see Figures
\ref{fig:fom2}, \ref{fig:fom7}, \ref{fig:jakobi}.

\subsubsection*{The atom $\mathbf{C}_{2}$.}

$\mathcal{L}$ consists of two orientable circles $\gamma_{1}$,
$\gamma_{2}$ and four heteroclinic $2$-dimensional separatrices
$\mathcal{S}_{1}$, $\mathcal{S}_{2}$, $\mathcal{S}_{3}$,
$\mathcal{S}_{4}$. Trajectories on $\mathcal{S}_{1}$,
$\mathcal{S}_{3}$ are approaching $\gamma_{1}$ as time tend to
$\infty$, and $\gamma_{2}$ as time tend to $-\infty$, while those
placed on $\mathcal{S}_{2}$, $\mathcal{S}_{4}$ approach
$\gamma_{2}$ as time tend to $\infty$, and $\gamma_{1}$ as time
tend to $-\infty$. Each of the sets
$\mathcal{L}_{\varepsilon}^{+}$, $\mathcal{L}_{\varepsilon}^{-}$
contains two families of Liouville tori. As $K$ approaches to $c$,
the tori
from one family in $\mathcal{L}_{\varepsilon}^{-}$ deform to $\mathcal{S}%
_{1}\cup\mathcal{S}_{2}$, and from the other one to $\mathcal{S}_{3}%
\cup\mathcal{S}_{4}$. The tori from one family in
$\mathcal{L}_{\varepsilon }^{+}$ is deformed to
$\mathcal{S}_{1}\cup\mathcal{S}_{4}$, and from the other to
$\mathcal{S}_{2}\cup\mathcal{S}_{3}$.

\subsubsection*{The atom $\mathbf{D}_{1}$.}

$\mathcal{L}$ consists of two orientable circles $\gamma_{1}$,
$\gamma_{2}$ and four $2$-dimensional separatrices
$\mathcal{S}_{1}$, $\mathcal{S}_{2}$, $\mathcal{S}_{3}$,
$\mathcal{S}_{4}$. Trajectories of $\mathcal{S}_{1}$,
$\mathcal{S}_{2}$ homoclinically tend to $\gamma_{1}$,
$\gamma_{2}$ respectively. Trajectories on $\mathcal{S}_{3}$ are
approaching $\gamma_{1}$ as time tend to $\infty$, and
$\gamma_{2}$ as time tend to $-\infty$, while those placed on
$\mathcal{S}_{4}$, approach $\gamma_{2}$ as time tend to $\infty$,
and $\gamma_{1}$ as time tend to $-\infty$. One of the sets
$\mathcal{L}_{\varepsilon}^{+}$, $\mathcal{L}_{\varepsilon}^{-}$
contains three, and the other one family of Liouville tori. Lets
say that $\mathcal{L}_{\varepsilon}^{+}$ contains three families.
As $K$ approaches to $c$, these families deform to
$\mathcal{S}_{1}$, $\mathcal{S}_{2}$ and
$\mathcal{S}_{3}\cap\mathcal{S}_{4}$ respectively, while the
family contained in $\mathcal{L}_{\varepsilon}^{-}$ deform to the
whole separatrix set.

\

This concludes the complete list of all atoms appearing in this
paper.

\subsection*{Numerical Fomenko Invariants}

Each edge of the Fomenko graph corresponds to a one-parameter
family of Liouville tori. Let us cut each of these families along
one Liouville torus. The manifold $\mathcal{Q}$ will disintegrate
into pieces, each corresponding to the singular level set, i.e. to
a part of the Fomenko graph containing only one vertex and the
initial segments of the edges incident to this vertex. To
reconstruct $\mathcal{Q}$ from these pieces, we need to identify
the corresponding boundary tori. This can be done in different
ways. Thus, the basis of cycles is fixed in a certain canonical
way on each boundary torus. Denote by $b_i^-$, $b_i^+$ the bases
corresponding to the beginning and the end of an edge
respectively. Denote by:
$$
\left(
\begin{array}{cc}
\alpha_i & \beta_i\\
\gamma_i & \delta_i
\end{array}
\right)
$$
the transformation matrix from $b_i^-$ to $b_i^+$.

\subsubsection*{Marks $r_{i}$}
They are defined as:
$$
r_i=
 \begin{cases}
 \left\{\dfrac{\alpha_i}{\beta_i}\right\}, &\text{if}\ \beta_i\neq0,
\\
\ \ \ \infty, &\text{if}\ \beta_i=0.
\end{cases}
$$

\subsubsection*{Marks $\varepsilon_{i}$}
They are:
$$
\varepsilon_i=
 \begin{cases}
\ \sign\beta_i, &\text{if}\ \beta_i\neq0,
\\
\ \sign\alpha_i, &\text{if}\ \beta_i=0.
\end{cases}
$$

\subsubsection*{Marks $n_k$}
To determine marks $n_k$, we cut all the edges of the graphs
having finite marks $r_i$. In this way the graph will fall apart
into a few connected parts. Consider only parts not containing
$\mathbf A$-atoms. To each edge of the graph having a vertex in a
given connected part, we join the following integer:
$$
\Theta_i=
\begin{cases}
\ \left[\dfrac{\alpha_i}{\beta_i}\right], &\text{if $e_i$ is has
the initial vertex in the connected part}
 \\
 &
 \\
\ \left[\dfrac{-\delta_i}{\beta_i}\right], &\text{if $e_i$ is has
the ending vertex in the connected part}
 \\
  &
 \\
\ \left[\dfrac{-\gamma_i}{\alpha_i}\right], &\text{if both
verteces of $e_i$ are in the connected part.}
\end{cases}
$$
Then to the connected part of the graph, we join the mark:
$$
n_k=\sum\Theta_i.
$$

\subsection*{Rotation Functions}

Consider an arbitrary edge of a given Fomenko graph and recall
that it represents a one-parameter family of Liouville tori. Let
us choose one such a torus and fix on it a basis $(\lambda,\mu)$
of cycles. By Arnol'd-Liouville theorem, the motion on the torus
is linear, thus there are coordinates
$(\varphi_1,\varphi_2)\in\mathbf S^1\times\mathbf S^1$ such that
the Hamiltonian vector field can be written in the form:
$$
a\frac{\partial}{\partial\varphi_1}+b\frac{\partial}{\partial\varphi_2},
$$
while the coordinate lines $\varphi_2=\const$, $\varphi_1=\const$
are equivalent to basic cycles $\lambda$, $\mu$. \emph{Rotation
number} on the torus is $\rho=\dfrac ab$. If we continuously
extend the basis $(\lambda,\mu)$ to the other tori of the family,
the collection of obtained rotation numbers will represent
\emph{rotation function}.

\bibliographystyle{amsalpha}

\end{document}